\pdfoutput=1
\RequirePackage{ifpdf}
\ifpdf 
\documentclass[pdftex]{sigma}
\else
\documentclass{sigma}
\fi

\usepackage{mathrsfs}
\usepackage[scr=boondoxo,scrscaled=1.05]{mathalfa}

\numberwithin{equation}{section}

\newtheorem{Theorem}{Theorem}[section]
\newtheorem*{Theorem*}{Theorem}
\newtheorem{Corollary}[Theorem]{Corollary}
\newtheorem{Lemma}[Theorem]{Lemma}
\newtheorem{Proposition}[Theorem]{Proposition}

\theoremstyle{definition}
\newtheorem{Definition}[Theorem]{Definition}

\newtheorem{Remark}[Theorem]{Remark}

\def\W{\mathscr{W} \mathrm{res}}
\def\Tr{\mathrm{Tr}}
\def\PDO{\Psi\mathrm{DO}}
\def\dv{\mathrm{dvol}_g}

\begin{document}

\allowdisplaybreaks

\newcommand{\arXivNumber}{2505.16642}

\renewcommand{\thefootnote}{}

\renewcommand{\PaperNumber}{035}

\FirstPageHeading

\ShortArticleName{On Geometric Spectral Functionals}

\ArticleName{On Geometric Spectral Functionals\footnote{This paper is a~contribution to the Special Issue on Asymptotics, Randomness and Noncommutativity. The~full collection is available at \href{https://sigma-journal.com/noncommutativity.html}{https://sigma-journal.com/noncommutativity.html}}}

\Author{Arkadiusz BOCHNIAK~$^{\rm ab}$, Ludwik D\c{A}BROWSKI~$^{\rm c}$, Andrzej SITARZ~$^{\rm d}$ and Pawe{\l} ZALECKI~$^{\rm d}$}

\AuthorNameForHeading{A.~Bochniak, L.~D\c{a}browski, A.~Sitarz and P.~Zalecki}

\Address{$^{\rm a)}$~Max-Planck-Institut f{\"u}r Quantenoptik, Hans-Kopfermann-Str.~1, Garching, 85748, Germany}

\Address{$^{\rm b)}$~Munich Center for Quantum Science and Technology,\\
\hphantom{$^{\rm b)}$}~Schellingstra{\ss}e~4, M{\"u}nchen, 80799, Germany}
\EmailD{\mail{arkadiusz.bochniak@mpq.mpg.de}}

\Address{$^{\rm c)}$~Scuola Internazionale Superiore di Studi Avanzati, Via Bonomea 265, Trieste, 34136, Italy}
\EmailD{\mail{dabrow@sissa.it}} %

\Address{$^{\rm d)}$~Institute of Theoretical Physics, Jagiellonian University,\\
\hphantom{$^{\rm d)}$}~{\L}ojasiewicza 11, Krak\'ow, 30-348, Poland}
\EmailD{\mail{andrzej.sitarz@uj.edu.pl}, \mail{pawel.zalecki@doctoral.uj.edu.pl}}

\ArticleDates{Received May 23, 2025, in final form March 24, 2026; Published online April 14, 2026}

\Abstract{We investigate spectral functionals associated with Dirac and Laplace-type differential operators on manifolds, defined via the Wodzicki residue, extending classical results for Dirac operators derived from the Levi-Civita connection to geometries with torsion. The local densities of these functionals recover fundamental geometric tensors, including the volume form, Riemannian metric, scalar curvature, Einstein tensor, and torsion tensor. Additionally, we introduce chiral spectral functionals using a grading operator, which yields novel spectral invariants. These constructions offer a richer spectral-geometric characterization of manifolds.}


\Classification{58B34; 58J50; 46L87; 83C65}

\renewcommand{\thefootnote}{\arabic{footnote}}
\setcounter{footnote}{0}

\section{Introduction}

Methods from Riemannian and Lorentzian geometry provide the mathematical foundation for modern theories of physical interactions. These differential geometric frameworks are central not only to Einstein's general theory of relativity -- which has been extensively validated through observations -- but also to a broad array of theoretical extensions and modifications that seek to address unresolved questions in cosmology and explore phenomena beyond the classical scope of general relativity. Differential operators defined on a smooth manifold play a fundamental role in the mathematical formulation of physical theories and are key elements in both quantum mechanics and quantum field theory. Their spectral properties determine the energy levels of systems and influence many physical phenomena.

From a mathematical perspective, the study of spectra of such operators -- such as the Laplace--Beltrami operator $\Delta$ or the Dirac operator $D$ -- has led to the emergence and rapid development of a vibrant branch of mathematics known as spectral geometry. This field investigates how the eigenvalues and eigenfunctions of differential operators reflect the geometric and topological features of the underlying space. One of the most iconic and influential questions in this context was popularized by Mark Kac: ``{\it Can one hear the shape of a drum?}'' \cite{Kac}. Though the answer is nuanced and context-dependent, it inspired a deeper exploration of the relationship between geometry and analysis.

This line of inquiry has since evolved into a powerful paradigm, suggesting that key properties of physical systems -- and even the structure of spacetime itself -- can be encoded in the spectrum of certain operators. In particular, within the framework of Alain Connes' noncommutative geometry \cite{Chamseddine:1996rw,Connes_book} and spectral geometry, this idea has been extended to propose that the dynamics and topology of spacetime may be recovered from the spectral data of appropriate differential operators. This approach offers a compelling bridge between abstract mathematical structures and physical reality.

In classical spectral geometry, the primary objects of study are spectral quantities defined for (pseudo-)differential operators acting on sections of vector bundles over smooth real manifolds. They typically depend on the asymptotics of the spectrum of these operators and often take the form of exotic traces, heat kernel coefficients, zeta functions, or determinant-like quantities.

When these operators act on sections of vector bundles -- such as the tangent bundle or the spinor bundle -- they reflect additional geometric and topological structures. For example, the heat trace asymptotics link the short-time behavior of the heat kernel to local geometric invariants like scalar curvature, while the spectral zeta functions encode global properties \cite{Gilkey, Gilkey2}.

An alternative and particularly elegant approach to defining spectral functionals makes use of the Wodzicki residue~$\W$ -- a powerful tool in the analysis of pseudo-differential operators. The Wodzicki residue is remarkable in that it provides, up to multiplication by a constant, the unique trace on the space of classical pseudodifferential operators acting on sections of a complex vector bundle over a compact, oriented manifold of dimension $n\geq 2$, when considered modulo smoothing operators and within fixed integer orders \cite{Guillemin,Wodzicki}. Rather than extending the usual operator trace, it defines a trace functional on a broader class of operators as the integral of a well-defined local density, thereby encoding geometric information and relating the local structure of an operator to global geometric invariants. The local densities appearing in spectral functionals of geometric significance correspond to geometric invariants of the underlying manifold or vector bundle, thereby justifying the nomenclature adopted for such functionals. These functionals can then be generalized to more abstract spectral triples within the framework of noncommutative geometry.

Let us briefly review the spectral functionals of particular physical interest. First, recall that the metric functional $\mathscr{g}_D$ and the Einstein functional $\mathscr{g}_D$ were constructed for spectral triples \cite[Definition~5.4]{dabrowski2023} and, respectively, assign to a pair of one-forms $(u,w)$ a number
\[
	\mathscr{g}_D(u,w)=\W(\hat{u}\hat{w}|D|^{-n}), \qquad u,w\in \Omega_D^1,
\]
and
\[
	\mathscr{G}_D(u,w)=\W\bigl(\hat{u}\{D,\hat{w}\} D |D|^{-n}\bigr), \qquad u,w\in \Omega_D^1,
\]
where $\hat{u}$ denotes the Clifford multiplication by the one-form $u\in \Omega^1_D$.

The torsion functional, as defined in \cite[Definition~2.1]{dabrowski2024}, assigns to a triple of one-forms $(u,v,w)$
\[
	\mathscr{T}_D(u,v,w)= \W\bigl(\hat{u}\hat{v}\hat{w}D D^{-2m}\bigr), \qquad u,v,w\in \Omega^1_D.
\]
Finally, the scalar curvature functional is defined by $\mathscr{R}_D(f)=\W\bigl(fD^{-2m+2}\bigr)$ for $f\in \mathcal{A}$ \cite{Ackermann96,Kalau95}. This functional, when regarded as a functional depending on the metric that determines the Dirac operator, is commonly known as the Einstein--Hilbert functional, since it directly corresponds to the Einstein--Hilbert action -- the cornerstone of general relativity.

The spectral functionals mentioned above have been extensively studied from multiple viewpoints \cite{dabrowski2023,dabrowski2024, dabrowski2024a}, particularly focusing on cases where the underlying geometric structure includes nontrivial torsion \cite{bochniak2025,dabrowski2025}, aiming to understand how it affects both the spectral invariants and the resulting physical models. This offers insights into possible generalizations of Einstein’s theory and alternative gravity frameworks. Some of these functionals have also been recently extended to manifolds with boundary; see \cite{Yang24} and the references therein.

In this paper, we provide a comprehensive study of the key properties of these functionals under the most general admissible perturbations of the Dirac operator, both in full generality and in specific cases, such as the spinorial and Hodge--Dirac settings.

\section{Spectral functionals: preliminaries}
\label{sec:general}
In this section, we first recall the results of computations of Wodzicki residues for a broad class of Laplace- and Dirac-type operators acting on sections of a vector bundle $V$ over a manifold~$M$. Here, $M$ always denotes a compact, closed Riemannian manifold of dimension $n = 2m > 1$ with a fixed metric.\footnote{We remark that most of the results presented here remain valid in the odd-dimensional case, as observed in~\cite{bochniak2025} for the spin-Dirac operator. For the sake of concreteness and generalizations involving gradings, we focus here on even-dimensional manifolds.}

We follow the conventions and notation used previously in \cite{dabrowski2023,bochniak2025}; for example, we work in normal coordinates at a fixed point and expand the homogeneous symbols of the operators under consideration in these coordinates. We recall the notion of a Laplace-type operator discussed in~\cite{bochniak2025}. An operator $L$ acting on $\Gamma(V)$, the module of smooth sections of $V$, is said to be of Laplace type if its symbol at any point $x \in M$ is given by $\sigma(L) = \mathfrak{a}_2 + \mathfrak{a}_1 + \mathfrak{a}_0$, where the homogeneous symbols $\mathfrak{a}_\bullet$, computed in normal coordinates $x$, are
\begin{gather}
		\mathfrak{a}_2= \left(\delta_{ab}+\frac{1}{3}R_{acbd}x^c x^d \right)\xi_a\xi_b + o\bigl({\bf x}^2\bigr),\qquad
		\mathfrak{a}_1={\rm i}\bigl(P_{ab}x^b + S_a\bigr)\xi_a +o({\bf x}),\nonumber\\
		\mathfrak{a}_0=Q+o({\bf 1}).\label{symbolL}
\end{gather}
Here, $R_{acbd}$ is the Riemann curvature tensor, and $P_{ab}$, $S_a$, and $Q$ are $\operatorname{End}(V)$-valued tensors evaluated at $x=0$. We recall that the Wodzicki residue \cite{Wodzicki} is (up to normalization) the unique trace on the algebra $\PDO(V)$ of classical pseudodifferential operators
\[
	\W(P)=\int_M\dv \int_{|\xi|=1} {\rm d}^4x \, \Tr(\sigma_{-n}(P)(x,\xi)),
\]
where $\sigma_{-n}(P)$ denotes the symbol of $P \in \PDO(V)$ of order $-n$ \cite{Gilkey}.

To compute the relevant functionals, we need the following general result.

\begin{Proposition}[{\cite[Proposition~I.2]{bochniak2025}}]
	\label{prop:main}
	For a second-order differential operator $O$, with a symbol expressed in normal coordinates around a point on $M$,
	\[\sigma(O)=F^{ab}\xi_a\xi_b + {\rm i} G^a \xi_a +H + o({\bf 1}),\]
	where $F^{ab} = F^{ba}$, $G^a$, and $H$ are endomorphisms of the fiber $V$ at the point $x = 0$, we have
	\begin{align*}
		\W(OL^{-m})={}&\frac{\nu_{n-1}}{24}\int_M \dv\Tr \bigl[24H +12 G^aS_a +F^{aa}(-12 Q+ 6P_{bb}-2 R - 3 S_b S_b) \\
		& \hphantom{\frac{\nu_{n-1}}{24}\int_M \dv\Tr \bigl[}{} + 2 F^{ab}(-6 P_{ab}+2 \mathrm{Ric}_{ab}-3S_a S_b)\bigr],
	\end{align*}
where $R$ is the scalar curvature and $\mathrm{Ric}$ is the Ricci tensor.
\end{Proposition}

We now present two corollaries that will be particularly useful.

\begin{Corollary}[{\cite[Corollary~I.3]{bochniak2025}}]
	\label{lemma:E}
	For a $C^\infty(M)$-endomorphism $E\colon S \to S$, we have
\[
	\W\bigl(EL^{-m+1}\bigr)=\frac{n-2}{24}\nu_{n-1}\int_M \dv \Tr[E(-12Q+6P_{aa}-2R-3S_aS_a)].
\]
\end{Corollary}

\begin{Corollary}[{\cite[Corollary~I.4]{bochniak2025}}] \label{cor4}
	For an operator $O$ with a symbol as in Proposition~{\rm\ref{prop:main}},
	we have
		\begin{align*}
			& \W \left( \left(O - \frac{1}{n-2}F^{aa}L\right) L^{-m} \right)\\
&\qquad{} = 	
			\frac{\nu_{n-1}}{24}\int_M \dv \Tr[24H +12 G^aS_a + 2F^{ab}(-6P_{ab}+2 \mathrm{Ric}_{ab}-3S_aS_b)].
		\end{align*}
\end{Corollary}

A particularly important class of Laplace-type operators arises from Dirac-type operators. We assume that $L = D^2$, where $D$ is a first-order elliptic differential operator acting on $\Gamma(V)$, with~$V$ being a Clifford module. We fix a reference Dirac operator $D_0$ and consider its perturbations~${D = D_0 + B}$, where $B$ is an endomorphism of the vector bundle. While the choice of $D_0$ and $B$ is not unique, in most cases there exists a preferred choice of~$D_0$, such as the standard spin Dirac operator (in the spin${}_c$ case) or the Hodge--de~Rham Dirac operator (in the oriented Riemannian case).

We impose minimal assumptions on $D_0$ ensuring that $D_0^2$ is of Laplace type.

\begin{Definition}\label{defi2}
	We say that $D = D_0 + B$ is a Dirac-type operator on sections of the vector bundle $V$ if the symbol of $D_0$ in normal coordinates around a fixed point on the manifold is given by $\sigma(D_0)=\mathfrak{d}_1+\mathfrak{d}_0$,
	where
	\[\mathfrak{d}_1=\bigl(-\gamma^a+f^a_{bc}x^b x^c\bigr)\xi_a+ o\bigl({\bf x}^2\bigr), \qquad \mathfrak{d}_0= g_a x^a+ o({\bf x}),\]
	for some smooth endomorphisms $f^a_{bc}$ (symmetric in $b$, $c$) and $g_a$ of $V$ evaluated at $x=0$. Additionally, the endomorphism $B$ has an expansion
	$B=B_0+B_a x^a +o({\bf x})$.
\end{Definition}

The structure of $\mathfrak{d}_1$ and $\mathfrak{d}_0$, together with the requirement that $\mathfrak{d}_1^2 = \mathfrak{a}_2$ as in \eqref{symbolL}, ensures that both $D_0^2$ and $D^2$ are Laplace-type. Notably, the absence of linear terms in ${\bf x}$ in the principal symbol $\mathfrak{d}_1$, and the vanishing of the zero-order symbol $\mathfrak{d}_0$ at $x=0$, are essential conditions\footnote{One could define Dirac-type operators without requiring the absence of linear terms in ${\bf x}$ in the principal symbol. However, for our purposes, this condition is essential.}. The explicit form of the symbols of $D_0$ (e.g., coefficients $f^a_{bc}$ and $g_a$) will not matter, as we focus only on the perturbation by $B$. The $\gamma^a$ are Clifford algebra generators on each fiber $V$ satisfying~${\bigl\{\gamma^a,\gamma^b\bigr\} = 2\delta^{ab}\mathbf{1}}$. Next, only the first two terms in the expansion of $B$ are relevant, since $B$ does not enter the principal symbol of $D$. Finally, since $D^2=D_0^2+\{D_0, B\}+B^2$, and using the relation $\{{\rm i}\gamma^a \partial_a, B\} = {\rm i}\{\gamma^a, B\} \partial_a + {\rm i}\gamma^a \partial_a B$, we get
	\begin{gather}
			 P_{ab} = P_{ab}^0 + {\rm i}\{\gamma^a, B_b\}, \qquad S_a = {\rm i}\{\gamma^a, B_0\}, \nonumber\\
			 S_a^0 = 0, \qquad Q = Q^0 + {\rm i}\gamma^a B_a + B_0^2,\label{eq:PSQ}
	\end{gather}
 with $P^0_{ab}={\rm i}\{\gamma^a,g_b\}+2\gamma^cf^a_{bc}$,
$S_a^0=0$, $ Q^0={\rm i}\gamma^ag_a$.
\begin{Proposition}
	\label{lemma:ED}
	Let $D$ be as in Definition {\rm\ref{defi2}}. Then, for $O = ED$, where $E$ is an arbitrary bundle endomorphism, we have
	\[
			\W\bigl(EDD^{-2m}\bigr)=
			\frac{\nu_{n-1}}{2}\int_M \dv \Tr\left[ E(2B_0-\gamma^a\{\gamma^a,B_0\}) \right].
		\]
\end{Proposition}

\begin{proof}
	Observe that $O = ED$ has the structure from Proposition~\ref{prop:main}, with $F^{ab} = 0$, $G^a = {\rm i}E\gamma^a$, and $H = EB_0$. The result follows directly from Proposition~\ref{prop:main}, with $L = D^2$, and the following observations: For $k > 0$, the leading symbols of $D^{-2k}$ read
$\sigma\bigl(D^{-2k}\bigr)=\mathfrak{c}_{2k}+\mathfrak{c}_{2k+1}+\mathfrak{c}_{2k+2}$,
where%
\begin{gather*}
		\mathfrak c_{2k} = \mathfrak c^0_{2k} + o\bigl({\bf x}^2\bigr),\qquad
		\mathfrak c_{2k+1}= \mathfrak c^0_{2k+1}+k \xi_a||\xi||^{-2k-2} \{\gamma^a,B\} + o({\bf x}),\\
		\mathfrak c_{2k+2}= \mathfrak c^0_{2k+2}-k||\xi||^{-2k-2} \bigl({\rm i}\gamma^aB_a+B^2_0\bigr) \\
		 \phantom{\mathfrak c_{2k+2}=}{} +k(k+1) ||\xi||^{-2k-4}\left({\rm i}\{\gamma^a,B_b\}+\frac12\bigl\{\gamma^a,B_0\bigr\}\bigl\{\gamma^b,B_0\bigr\}\right) \xi_a \xi_b + o({\bf 1}).
\end{gather*}
Here, $\mathfrak{c}_p^0$ denotes the respective symbols of $D_0^{-2k}$
$\sigma\bigl(D_0^{-2k}\bigr) = \mathfrak{c}_{2k}^0 + \mathfrak{c}_{2k+1}^0 + \mathfrak{c}_{2k+2}^0$. This, together with \eqref{eq:PSQ}, concludes the claim.
\end{proof}

From now on, we assume that the Dirac operator fulfills the above requirements.

\section{General results on spectral functionals}

We begin this section with a result for a specific type of perturbation of $B$ of a Dirac operator.
\begin{Proposition}
	\label{rem:0}
	If $B= b_a\gamma^a$, where $b_a$ are endomorphisms commuting with the Clifford module $($in particular, $B$ can be an image of a one-form in the Clifford algebra$)$, then for every endomorphism $E$, $\W\bigl(EDD^{-2m}\bigr) = 0$, so that $D$ is spectrally closed, that is for any zero-order operator $O$ the Wodzicki residuum of $OD|D|^{-2m}$ vanishes $($cf. {\rm\cite[\emph{Definition}~5.5]{dabrowski2023})}.
\end{Proposition}
\begin{proof}
	To demonstrate this, we compute the density depending on $B$ using Proposition~\ref{lemma:ED}
	\[
		\Tr\bigl[E \bigl(2 b_b\gamma^b - \gamma^a
		\{\gamma^a,b_c\gamma^c\}\bigr)\bigr]
		=\Tr\bigl( E \bigl( 2 b_b \gamma^b -
		b_c \gamma^a 2 \delta^{ac} \bigr) \bigr) =0.\tag*{\qed}
	\]\renewcommand{\qed}{}
\end{proof}

\begin{Proposition}
	\label{rem:1}
	If $E=e_a\gamma^a$, where $e_a$ are endomorphisms commuting with the Clifford module $($in particular, $E$ can be an image of a one-form in the Clifford algebra$)$, then the functional~$\W\bigl(EDD^{-2m}\bigr)$ vanishes identically for every $B$.
\end{Proposition}
\begin{proof}
	This follows from the following computations:
		\begin{align*}
			\Tr\bigl[e_b\gamma^b(2B_0-\gamma^a\{\gamma^a,B_0\})\bigr]&=\Tr \bigl[e_b\bigl(2\gamma^bB_0-\gamma^b\gamma^a\gamma^aB_0-\gamma^b\gamma^aB_0\gamma^a\bigr)\bigr]\\
			&=\Tr \bigl[ e_b\bigl(2\gamma^bB_0-\bigl\{\gamma^b,\gamma^a\bigr\}\gamma^aB_0\bigr)\bigr]=0.\tag*{\qed}
		\end{align*}\renewcommand{\qed}{}
\end{proof}

We remark that in Proposition~\ref{rem:0}, we assume the form of the endomorphism $B$ and compute the functional $\W\bigl(EDD^{-2m}\bigr)$ for arbitrary endomorphism $E$, while in Proposition~\ref{rem:1}, the form of $E$ is assumed and the result is stated for arbitrary $B$.

Next, we study the dependence of the spectral functionals on the perturbations $B$ of the Dirac operator $D_0$. We have the following.

\begin{Proposition}[{cf.\ \cite[Theorem~4.1]{dabrowski2023}}]
	\label{prop:met}
	The metric functional does not depend on any bounded perturbation $B$,
	\[
		\mathscr{g}_D(u,w) = \mathscr{g}_{D_0}(u,w) = \mathrm{dim}(V)
		\nu_{n-1} \int_M \dv g(u,v).
	\]
\end{Proposition}
We emphasize that the above result is more general than in \cite[Theorem~4.1]{dabrowski2023} and shows that, regardless of the specific form of $D$, it depends only on the metric $g$ on $M$. The proof follows directly from Proposition~\ref{prop:main} with $F^{ab}=G^c=0$ and $H=\hat{u}\hat{v}$.

A more interesting case is the Einstein functional.
\begin{Proposition}
	\label{prop:Ein}
	For $D=D_0+B$ satisfying the assumptions of this paper $($cf.\ Definition~{\rm\ref{defi2})}, the Einstein functional density in normal coordinates reads
		\begin{align*}
			\mathscr{G}_D(u,w)(x)={} &\mathscr{G}_{D_0}(u,w)(x)
			+\frac{\nu_{n-1}}{2} 
			\Tr\biggl\{ iu_aw_{bc}\big[\gamma^a,\gamma^b\big]\{\gamma^c, B_0\}+\frac{\rm i}{2}u_aw_{b}\bigl[\gamma^a,\gamma^b\bigr]\{\gamma^c, B_c\} \\
			& +u_aw_b\bigl[\bigl(\delta^{ab}B_0-\gamma^a\bigl\{\gamma^b,B_0\bigr\}\bigr)(\{\gamma^c,B_0\}\gamma^c-2B_0)\bigr]\biggr\}.
		\end{align*}
\end{Proposition}
\begin{proof}
	The proof is identical to the first part in \cite[Proposition~II.1]{bochniak2025}.
\end{proof}

\begin{Corollary} \label{example:fl}
	An immediate application of the above result is that the Einstein functional is invariant under fluctuations of the Dirac operator $D$ of the form $A_a \gamma^a$, where $A_a$ are endomorphisms commuting with the Clifford module.
\end{Corollary}
\begin{proof}
	For $B'=B+A_a\gamma^a$, we have $\{\gamma^c, B'-B\}=2A_a\delta^{ac}$, therefore $\{\gamma^c, B_0'-B_0\}\gamma^c - 2(B_0'-B_0)=0$ and $\Tr\bigl[\gamma^a,\gamma^b\bigr]\{\gamma^c, B'-B \}=0$. The last identity ensures that the first two linear terms in $B'$ remain as they were for $B$. To complete the proof, it remains to show that the quadratic terms are the same for $B$ and $B'=B+A_a\gamma^a$. Indeed, $\{\gamma^c, B_0'\}\gamma^c -2B_0'=\{\gamma^c, B_0\}\gamma^c - 2B_0$, while
	\[
	\bigl(\delta^{ab}B_0'-\gamma^a\bigl\{\gamma^b,B_0'\bigr\}\bigr)-\bigl(\delta^{ab}B_0 -\gamma^a \bigl\{\gamma^b, B_0\bigr\}\bigr)=\gamma^c \bigl(\delta^{ab} A_c -2\delta^{ac} A_b\bigr).
	\]
	The claim follows from the proof of Proposition~\ref{rem:1}.
\end{proof}

Next, we discuss how a perturbation $B$ affects the torsion functional.
\begin{Proposition}\label{prop:ogolna_torsja}
	For the Dirac-type operator $D=D_0+B$, we have
	\begin{align*}
			\mathscr{T}_D(u,v,w)&=
			\nu_{n-1}\int_M \dv u_av_bw_c\Tr\bigl[\bigl(\gamma^a\gamma^c\gamma^b-\gamma^b\gamma^c\gamma^a\bigr)B_0\bigr]\\
			&=\frac{\nu_{n-1}}{2}\int_M \dv u_av_bw_c\Tr\bigl(\bigl[\gamma^b,\gamma^a\bigr]\{\gamma^c,B_0\}\bigr).
		\end{align*}
	The result is totally antisymmetric in $u$, $v$, $w$.
\end{Proposition}
\begin{proof}
	First, note that the torsion functional vanishes for $D_0$. The result is a direct consequence of Corollary~\ref{lemma:ED} for $E=u_av_bw_c\gamma^a\gamma^b\gamma^c$. We obtain
		\begin{align*}
			\mathscr{T}_D(u,v,w)={}&\frac{\nu_{n-1}}{2}\int_M \dv u_av_bw_c\Tr\bigl[\gamma^a\gamma^b\gamma^c(2B_0-\gamma^d\bigl\{\gamma^d,B_0\bigr\})\bigr]\\
			={}&\frac{\nu_{n-1}}{2}\int_M \dv u_av_bw_c\Tr\bigl[\bigl(2\gamma^a\gamma^b\gamma^c-\gamma^a\gamma^b\gamma^c\gamma^d\gamma^d-\gamma^d\gamma^a\gamma^b\gamma^c\gamma^d\bigr)B_0\bigr]\\
			={}&\frac{\nu_{n-1}}{2}\int_M \dv u_av_bw_c\Tr\\
&\times\bigl[\bigl(2\gamma^a\gamma^b\gamma^c-2\delta^{ad}\gamma^b\gamma^c\gamma^d+2\delta^{bd}\gamma^a\gamma^c\gamma^d-2\delta^{cd}\gamma^a\gamma^b\gamma^c\bigr)B_0\bigr]\\
			={}&\frac{\nu_{n-1}}{2}\int_M \dv u_av_bw_c\Tr\bigl[2\bigl(\gamma^a\gamma^c\gamma^b-\gamma^b\gamma^c\gamma^a\bigr)B_0\bigr].
		\end{align*}
	The combination $\gamma^a\gamma^c\gamma^b-\gamma^b\gamma^c\gamma^a$ is totally antisymmetric in $a$, $b$, $c$, which is clear for the pair~$a$,~$b$. For other pairs, e.g., $a$, $c$, we can write
	\[
		\gamma^c\gamma^a\gamma^b-\gamma^b\gamma^a\gamma^c
=\bigl(2\delta^{ac}\gamma^b-\gamma^a\gamma^c\gamma^b\bigr)-\bigl(2\gamma^b\delta^{ac}-\gamma^b\gamma^c\gamma^a\bigr)=-\bigl(\gamma^a\gamma^c\gamma^b-\gamma^b\gamma^c\gamma^a\bigr).
	\]
	The last form (with $\bigl[\gamma^b,\gamma^a\bigr]$) follows from splitting
	\[
		\gamma^a\gamma^c\gamma^b=\frac12\bigl(2\delta^{ac}\gamma^b-\gamma^c\gamma^a\gamma^b+2\delta^{bc}\gamma^a-\gamma^a\gamma^b\gamma^c\bigr),
	\]
	and similarly for $\gamma^b\gamma^c\gamma^a$, then using the trace property.
\end{proof}

\begin{Corollary}
	The torsion functional does not depend on fluctuations of the Dirac operator of the form $A_a \gamma^a$, where $A_a$ are endomorphisms commuting with the Clifford module.
\end{Corollary}

Finally, we study the impact of $B$ on the scalar curvature functional. Recall that the dependence of this functional on torsion has been extensively studied in some special cases, where torsion enters through Dirac operators and contributes additional terms to heat-kernel coefficients and the spectral action \cite{Iochum12,Pfaffle12}; however, no general formula applicable, for instance, to the Hodge--Dirac operator has been obtained so far.
	\begin{Proposition}
		\label{prop:SA}
		The scalar curvature spectral action functional for the Dirac-type operator, as in Proposition~{\rm\ref{lemma:ED}}, reads
		\[
				\mathscr{R}_D(f)=\mathscr{R}_{D_0}(f)+ \frac{(n-2)\nu_{n-1}}{24}\int_M \dv f \Tr\bigl(
				-12B_0^2 + 6\gamma^a\{\gamma^a,B_0\}B_0\bigr).
			\]
	\end{Proposition}
	
	\begin{proof}
		To compute the spectral functional, we start by applying Corollary~\ref{lemma:E} to an endomorphism $E=f$, i.e., a function on $M$,
		\[
			\mathscr{R}_D(f)=\frac{(n-2)\nu_{n-1}}{24}\int_M \dv f \Tr(-12Q+6P_{aa}-2R-3S_aS_a).
		\]
		Since the operator $D$ is determined by \eqref{eq:PSQ}, we have
		\begin{gather*}
				\Tr(-12Q+6P_{aa}-2R-3S_aS_a)\\
\qquad= \Tr\bigl(-12Q^0+6P^0_{aa}-2R-12B_0^2-12i\gamma^aB_a
				+6i\{\gamma^a,B_a\}+3\{\gamma^a,B_0\}\{\gamma^a,B_0\}\bigr)\\
				\qquad=\Tr\bigl(-12Q^0+6P^0_{aa}-2R-12B_0^2+3\{\gamma^a,B_0\}\{\gamma^a,B_0\}\bigr),
			\end{gather*}
		where in the last equality, we use the fact that $\Tr(\{\gamma^a,B_a\})=\Tr(2\gamma^aB_a)$. To complete the proof, we note that
		\[
				\Tr(\{\gamma^a,B_0\}\{\gamma^a,B_0\})=\Tr (\{\gamma^a,B_0\}\gamma^aB_0+\{\gamma^a,B_0\}B_0\gamma^a) =\Tr(2\gamma^a\{\gamma^a,B_0\}B_0),
			\]
		which follows from the fact that $\{\gamma^a,B_0\}\gamma^a=\gamma^a\{\gamma^a,B_0\}$ and the cyclicity of the trace.
		Note that the final result does not depend on the derivatives of $B$. The part of the expression that does not depend on $B$ yields the value of the scalar curvature functional for $L=D_0^2$.
	\end{proof}
	
	We conclude this subsection by computing the functional scalar curvature of any Dirac-type operator $D_0$ of the type assumed in Definition~\ref{defi2}.
	\begin{Corollary}
		For every $D_0$ that satisfies the assumptions in Definition {\rm\ref{defi2}}, we have
		\[ \mathscr{R}_{D_0}(f) = \frac{(n-2)\nu_{n-1}}{24}\int_M \dv (-R). \]
	\end{Corollary}
	
	\begin{proof}
		Recall that for $D^2_0$, we have (cf.\ Definition~\ref{defi2})
		\[
			P^0_{ab}={\rm i}\{\gamma^a,g_b\}+2\gamma^cf^a_{bc},
			\qquad S_a^0=0,\qquad Q^0={\rm i}\gamma^ag_a,
		\]
		and therefore
		\begin{align*}
			\Tr\bigl(-12Q^0+6P^0_{aa}-2R-3S^0_aS^0_a\bigr)&=
			\Tr(-12{\rm i}\gamma^ag_a+6{\rm i}\{\gamma^a,g_a\}+12\gamma^cf^a_{ac}-2R)\\
			&=\Tr(6\{\gamma^c,f^a_{ac}\}-2R),
		\end{align*}
		where terms containing $g_a$ cancel each other due to the property of the trace. Next, because of the symmetry of $f^a_{ac}$, we can write
		\[
		6\{\gamma^c,f^a_{ac}\}=3(\{\gamma^c,f^a_{ac}\}+\{\gamma^a,f^c_{ac}\})=-R_{caac}-R_{ccaa}=R,
		\]
		where we have used
		\[
			\left(\bigl\{\gamma^a,f^b_{cd}\bigr\}+\frac13 R_{acbd} \right)\xi_a\xi_b x^c x^d=0.
		\]
		This comes from the requirement that $D_0^2$ has the same expansion of the principal symbol as the Laplace operator.
		Substituting this into the previous equation yields the result.
	\end{proof}
	
\subsection{The spin Dirac operator}
As this case was the subject of a separate study \cite{bochniak2025}, we start by briefly summarizing
the main results concerning spectral functionals discussed therein, and slightly extend some of them.

We recall that the (torsion-less) spin Dirac operator is given by
\[
D_0={\rm i} \gamma^i\nabla_{e_i}^{(s)}={\rm i}\gamma^i e_i -\frac{1}{4}\omega_{ijk}\gamma^i\gamma^j\gamma^k,
\]
with the spin connection $\omega_{ijk}=\frac{1}{2}(c_{ijk}+c_{kij}+c_{kji})$ defined by structure constants $[e_i,e_j]=c_{ijk}e_k$. Since in the normal coordinates
\[
c_{pqr}=\frac{1}{2}R_{pqnr}x^n +o({\bf x}), \qquad
\omega_{ijk}=-\frac{1}{2}R_{nijk}x^n +o({\bf x}), \qquad
\omega_{ijk}\gamma^i\gamma^j\gamma^k
=\mathrm{Ric}_{ab}\gamma^b x^a,
\]
we have
\begin{equation}
	D_0={\rm i}\gamma^a\left(\partial_a - \frac{1}{6}R_{abcd}x^b x^c \partial_d-\frac{1}{4}\mathrm{Ric}_{ab} x^b\right),
	\label{Dirac}
\end{equation}
and therefore
\begin{align*}
D_0^2
=&-\left(\delta_{ab}+\frac{1}{3}R_{acbd}x^c x^d +o\bigl({\bf x}^2\bigr)\right)\partial_a\partial_b+\left[\frac{1}{4}R_{kjba}\gamma^j\gamma^k +\frac{2}{3}\mathrm{Ric}_{ab}\right]x^b \partial_a + \frac{1}{4}R.
\end{align*}
Our first application is the perturbation of $D_0$ by torsion, extending the results obtained in \cite{dabrowski2024}.
\begin{Proposition}
\label{example:t}
On a spin manifold $M$ with the spinor bundle $S$ and $D_0$ the Dirac operator~{\rm\eqref{Dirac}}, if the perturbation
$B$ is of the form
\begin{equation}
\label{eq:B}
B=-\frac{\rm i}{8}{}^AT_{ijk}\gamma^i\gamma^j\gamma^k,
\end{equation}
with a totally antisymmetric tensor ${}^AT_{ijk}$, to which we refer as a torsion, we have that the density of the Einstein functional in normal coordinates is
\begin{align*}
\mathscr{G}_D(u,w)(x)-\mathscr{G}_{D_0}(u,w)(x)
		={} &3 \cdot 2^{n-1} \nu_{n-1}\biggl[-u_aw_{bc}{}^AT^{0}_{abc} \\
			& +\frac{1}{8}u_aw_b\bigl(\delta^{ab}{}^AT^{0}_{ijk}{}^AT^{0}_{ijk}-4{}^AT^{c}_{abc}-6 {}^AT^{0}_{ajk}{}^AT^{0}_{bjk}\bigr)\biggr],
		\end{align*}
 we use the expansion ${}^AT_{ijk}={}^AT^{0}_{ijk}+{}^AT^{c}_{ijk}x^c +o({\bf x})$.
\end{Proposition}
\begin{proof}
By straightforward computations, we see that
	\begin{gather}
			\{\gamma^a, B\}=-\frac{3{\rm i}}{4}{}^AT_{ajk}\gamma^j\gamma^k, \qquad \gamma^a\{\gamma^a,B\}=\{\gamma^a, B\}\gamma^a=6B,\nonumber\\
			\Tr(\gamma^a B)=0, \qquad \Tr\bigl(\gamma^a\gamma^b\{\gamma^c, B\}\bigr)=2 \Tr\bigl(\gamma^a \gamma^b\gamma^c B\bigr)=-2^{m} \frac{3{\rm i}}{2}{}^AT_{cba}.\label{eq:slady_B_spinowe}
	\end{gather}
Using then Proposition~\ref{prop:Ein}, we end up with the form of the Einstein functional in the presence of torsion.
\end{proof}

Note that the above functional constitutes a nontrivial contribution to the Einstein functional arising from the antisymmetric torsion and that the resulting density is not tensorial (the density of the functional is not $C^\infty(M)$-bilinear), as it depends on the derivatives of the forms. Furthermore, we note that the part of the functional that is non-tensorial remains symmetric in~$u$,~$w$ up to a total derivative,
\[
 - u_a w_{bc}{}^AT^{0}_{abc} - \frac{1}{2} u_a w_{b} {}^AT^{c}_{abc}
= - \partial_c \bigl( u_a w_{b}{}^AT_{abc} \bigr)
 - w_a u_{bc}{}^AT^{0}_{abc} - \frac{1}{2} w_a u_{b} {}^AT^{c}_{abc}.
\]
In \cite{bochniak2025}, by considering a broader class of tensor-type functionals, we argue that the presence of such non-tensorial terms indicates an obstruction to the inclusion of torsion in physically acceptable models. It is also well known that lifting a connection with torsion to the spinor bundle may result in a non-self-adjoint operator unless the torsion is fully antisymmetric. Therefore, we can also consider non-self-adjoint bounded perturbations of the Dirac operator $D_0$. A detailed analysis of this class of operators will be conducted elsewhere~\cite{to_appear}.

Finally, we easily recover the result from \cite{dabrowski2024} for the torsion functional alone and the scalar curvature functional.
\begin{Proposition}[{cf.\ \cite[Theorem~2.2]{dabrowski2024}}]
	For the Dirac-type operator $D=D_0+B$ with $B$ given by equation~\eqref{eq:B}, the torsion functional reads
	\[
		\mathscr{T}_D(u,v,w)= -\frac{3{\rm i}}{2}\nu_{n-1}\dim(V)\int_M \dv u_{a}v_bw_c {}^AT^{0}_{abc}.
	\]
\end{Proposition}
\begin{proof}
It follows directly from Proposition~\ref{prop:ogolna_torsja} and by using \eqref{eq:slady_B_spinowe}.
\end{proof}

\begin{Theorem}\label{EH-spinD}
The scalar curvature functional for the torsion perturbation of the spin Dirac operator,
\smash{$B=-\frac{\rm i}{8}{}^AT_{abc}$} reads
\[
	\mathscr{R}_D(f)=\frac{2^m(n-2)\nu_{n-1}}{24}\int_M \dv f \left[-R+\frac{9}{4}{}^AT_{abc}^0{}^AT_{abc}^0\right].
\]
\end{Theorem}
\begin{proof}
This follows directly from
\[
	\Tr\bigl(-12 B_0^2 +6\gamma^a\{\gamma^a, B_0\}B_0\bigr)=2^m \cdot \frac{9}{4} {}^AT_{abc}^0 {}^AT_{abc}^0.\tag*{\qed}
\]\renewcommand{\qed}{}
\end{proof}

\begin{Remark}
The result presented in Theorem~\ref{EH-spinD} is consistent with those found in the existing literature, for example, in~\cite[p.~879]{hanisch10} (noting that their definition of $\bigl(T^0\bigr)^2$ differs from ours by a factor of $\frac{1}{6}$), as well as in~\cite[Proposition~5.4]{Pfaffle12} and~\cite[Section~5.2]{Kalau95}. We note that in the latter works, the term ``torsion'' refers to what is technically the contorsion, which introduces an additional factor of $\frac{1}{4}$ compared to our formulation.
However, there is an inconsistency with the formula given in~\cite[Lemma~3.1]{Ackermann96}, where a factor related to the rank of the corresponding bundle appears, which should not be present.
\end{Remark}

\subsection{The Hodge--Dirac operator}
We recall here the construction of the Hodge--Dirac operator from \cite{dabrowski2024a}; however, we introduce a~family of Dirac-type operators extended by torsion.

We review the basics of our notation from \cite{dabrowski2024a}. We use multi-index notation to represent differential forms and operations on them compactly. For $0\leq l\leq n$, a differential $l$-form $\omega$ is denoted by $\omega=\sum_{J} \omega_J {\rm d}x^J$, where $J$ is an ordered multi-index. Raising and lowering the degrees of differential forms are performed using the operators $\lambda^p_+$ and $\lambda_-^p$, respectively, whose components are given by
\[
 \bigl(\lambda^p_+\bigr)^I_J=\varepsilon^I_{pJ}, \qquad \bigl(\lambda^p_-\bigr)^I_J=\varepsilon^{pI}_J.
\]
They act on basis forms $e^J$ as
 \smash{$
 \lambda^p_+e^J=\varepsilon^I_{pJ}e^I$},
 \smash{$\lambda^p_-e^J=\varepsilon^{pI}_Je^I$},
where
\[
 \varepsilon^{I}_{pJ}=\begin{cases}
 (-1)^{\mathrm{sgn}(\pi)}, & \exists \mbox{ permutation } \pi \mbox{ s.t. } pJ=\pi(I),\\
 0, & \mbox{otherwise}.
 \end{cases}
\]
They satisfy
$
 \bigl\{\lambda_\pm^p,\lambda_\pm^r\bigr\}=0$, $ \bigl\{\lambda^p_+,\lambda^r_-\bigr\}=\delta_{pr}\mathrm{id}$,
and the operators $\gamma^p=-i\bigl(\lambda^p_+-\lambda^p_-\bigr)$ generates Clifford algebra $\{\gamma^p,\gamma^r\}=\delta_{pr}\mathrm{id}$.

The Hodge--Dirac operator, $D_0=d+d^\ast$, was studied from the spectral perspective in \cite{dabrowski2024a}. Here we study
the perturbed Hodge--Dirac operator, $D=D_0+B$ with
\begin{align}
 B&=\frac{1}{2} T_{ijk}\bigl(\lambda_+^j\lambda_+^i\lambda_-^k+\lambda_+^k\lambda_-^i\lambda_-^j\bigr) =\frac{1}{2}T_{ijk}\bigl[\delta_{ik}\bigl(\lambda^j_++\lambda^j_-\bigr)-\lambda^j_+\lambda^k_-\lambda^i_+ + \lambda^j_-\lambda^k_+ \lambda^i_-\bigr].\label{eq:HDB}
\end{align}
First, we demonstrate that the above $B$ can be considered as a generalization of the Hodge--Dirac operator with torsion. Since the exterior derivative $d$ is expressed using the Levi-Civita
connection ${\rm d} = \sum_i {\rm d}x^i \wedge \nabla^{\mathrm{LC}}_i$, we construct
a first-order differential operator $\tilde{\rm d} = \sum_i {\rm d}x^i \wedge \nabla_i$ for an arbitrary metric compatible linear connection $\nabla$. Explicitly, on a one-form $u$,
\[
 \bigl(\tilde{\rm d} u\bigr)_J=\bigl(\partial_j u_I-\Gamma^k_{jk} u_I + \varepsilon^{iK}_{kI} \Gamma^k_{ji}u_K\bigr)\varepsilon^{jI}_J=\bigl(\partial_j u_I-\varepsilon^{iA}_I\varepsilon^{K}_{kA} \Gamma^k_{ji}u_K\bigr)\varepsilon^{jI}_J,
\]
where $\Gamma^{c}_{ab}$ are the Christoffel symbols of the connection $\nabla$. Since $T_{ijk}=\Gamma^{k}_{ij}-\Gamma^{k}_{ji}$, the above expression can be equivalently rewritten as
\[
 \tilde{\rm d} u=\lambda_+^j \partial_j u + \frac{1}{2}\lambda_+^j \lambda_+^i\lambda_-^k T_{ijk}u.
\]
The Hodge--Dirac operator with torsion is then defined as $D=\tilde{\rm d}+\tilde{\rm d}^\ast$, which for $T_{ijk}=0$ is the usual
Hodge--Dirac $d+d^*$. Using the definition of an adjoint operator $A^\ast$ of an operator $A$, in normal coordinates we get
$
 (A^\ast)_{LJ}=\overline{A_{JL}} + o({\bf x})$.
Therefore,
\[
 \tilde{\rm d}^\ast=d^\ast+\frac{1}{2}\lambda_+^k \lambda_-^i \lambda_-^j T_{ijk}.
\]
Observe that the operator $D$ is a Dirac-type operator as defined in Definition~\ref{defi2}; thus, it yields the standard metric functional on $M$ (up to a multiplicative constant). We also note that a~Hodge--Dirac operator coupled to antisymmetric torsion appears in \cite{wang2024}; however, comparing it to the construction presented here is not straightforward.

\begin{Proposition}\label{prop:Ein_HD}
For the Hodge--Dirac operator with torsion \eqref{eq:HDB}, the Einstein functional densities at point $x$ in normal coordinates reads
\begin{align*}
\mathscr{G}_D(u,w)(x)={}&\mathscr{G}_{D_0}(u,w)(x)
			+ 3\nu_{n-1}2^{n-3} u_a\left[-4w_{bc}{}^AT^{0}_{abc}-2w_b {}^AT^{c}_{abc}\right. \\
 &\left.-\frac{4}{3}w_a T^0_{jii}T^{0}_{jkk} +w_b {}^AT^{0}_{cjk} {}^AT^{0}_{djk}\left(\frac{1}{2}\delta_{ab}\delta_{cd}-3\delta_{ac}\delta_{bd}\right) \right],
		\end{align*}
where ${}^AT_{ijk}=\frac{1}{3}(T_{ijk}+T_{kij}+T_{jki})$ and $\mathscr{G}_{D_0}(u,w)$ is given by {\rm\cite[\emph{Proposition}~3.3]{dabrowski2024}}.
\end{Proposition}
To prove it, we shall need some useful identities involving the trace comprised in the following lemmas.
\begin{Lemma}
	The following identities hold:
	\begin{gather*}
				 \Tr\bigl(\gamma^a \lambda^i_{\pm}\bigr)=\pm{\rm i} 2^{n-1}\delta_{ai}, \qquad
			 \Tr\bigl(\gamma^a \lambda^j_{\mp}\lambda^k_{\pm}\lambda^i_{\mp}\bigr)=\mp{\rm i} 2^{n-2}(\delta_{ai}\delta_{kj}+\delta_{aj}\delta_{ki}), \\
			 \Tr\bigl(\gamma^a\gamma^b \gamma^c \lambda^i_{\pm}\bigr)=\pm{\rm i} 2^{n-1}(\delta_{ai}\delta_{bc}-\delta_{bi}\delta_{ac}+\delta_{ci}\delta_{ab}), \\
				\Tr\bigl(\gamma^a\gamma^b\gamma^c \lambda^j_{\pm}\lambda^k_{\mp}\lambda^i_{\pm}\bigr)=\pm 2^{n-3}\bigl[2\delta_{bc}(\delta_{ai}\delta_{kj}+\delta_{aj}\delta_{ki})-2\delta_{ac}(\delta_{bi}\delta_{kj}+\delta_{bj}\delta_{ki})\\
		\phantom{\Tr\bigl(\gamma^a\gamma^b\gamma^c \lambda^j_{\pm}\lambda^k_{\mp}\lambda^i_{\pm}\bigr)=}{}
+2\delta_{ab}(\delta_{ic}\delta_{kj}+\delta_{cj}\delta_{ki})+\delta_{ka}\varepsilon^{ji}_{bc}+\delta_{kb}\varepsilon^{ji}_{ca}+\delta_{kc}\varepsilon^{ji}_{ab}\bigr],
 \\
			 \Tr\bigl(\gamma^a\gamma^b \lambda^j_{\mp}\lambda^i_{\mp}\bigr)=-2^{n-2}\varepsilon^{ij}_{ab}, \\
			 \Tr\bigl(\gamma^a\gamma^b\gamma^c\gamma^d \lambda^j_{\pm}\lambda^i_{\pm}\bigr)=2^{n-2}\bigl(\delta_{ab}\varepsilon^{ji}_{cd}+\delta_{ac}\varepsilon^{ji}_{db}+\delta_{ad}\varepsilon^{ji}_{bc}
+\delta_{bc}\varepsilon^{ji}_{ad}+\delta_{bd}\varepsilon^{ji}_{ca}+\delta_{cd}\varepsilon^{ji}_{ab}\bigr).	
	\end{gather*}
\end{Lemma}
Furthermore, by straightforward computations, we have the following.
\begin{Lemma}
	\label{lemma:TRB}
	With $B$ given by \eqref{eq:HDB} the following identities hold:
	\begin{gather*}
			 \Tr(\gamma^aB)=0, \qquad
			 \Tr\bigl(\gamma^a\gamma^b\gamma^cB\bigr)={\rm i} 2^{n-2}(T_{abc}+T_{cab}+T_{bca}),\\
			\Tr\bigl(\gamma^a\gamma^b\{\gamma^c,B\}\bigr)={\rm i} 2^{n-1}(T_{abc}+T_{cab}+T_{bca}),\\
			 \Tr\bigl(\gamma^a\gamma^b\gamma^c\gamma^d\bigl\{\gamma^d,B\bigr\}\bigr)=3{\rm i}\cdot 2^{n-1} (T_{abc}+T_{cab}+T_{bca}),\\
			 \Tr\bigl(B^2\bigr)=2^{n-2}T_{baa}T_{bcc}+2^{n-3}T_{abc}T_{abc},\\
			 \Tr\bigl(\gamma^a\bigl\{\gamma^b,B\bigr\}B\bigr)=
			-2^{n-2}T_{ajk}T_{bkj}+2^{n-2}T_{ajk}T_{bjk}+2^{n-3}T_{jka}T_{jkb}, \\
			\Tr\bigl(\gamma^c\gamma^a\bigl\{\gamma^b,B\bigr\}\{\gamma^c,B\}\bigr)=2^{n-1}[T_{jkb}(T_{jka}+T_{ajk})+T_{bkj}(T_{kja}+2T_{akj}-2T_{ajk})].	
	\end{gather*}
\end{Lemma}
\begin{proof}[Proof of Proposition~\ref{prop:Ein_HD}]
It follows from combining Proposition~\ref{prop:Ein} and Lemma~\ref{lemma:TRB}.
\end{proof}

\begin{Proposition}
 For the Hodge--Dirac operator with torsion \eqref{eq:HDB}, the torsion functional reads
 \[
 \mathscr{T}_D(u,v,w)= - 3 {\rm i}\nu_{n-1}2^{n-1}\int_M \dv u_{a}v_bw_c \bigl({}^AT^{0}_{abc}\bigr).
 \]
\end{Proposition}
\begin{proof}
 It is an immediate consequence of Lemma~\ref{lemma:TRB}.
\end{proof}

\begin{Remark}
Observe that the torsion functional detects only the antisymmetric part of the torsion.
\end{Remark}
\begin{Proposition}
 \label{prop:HD_SA}
 For the Hodge--Dirac operator with torsion~\eqref{eq:HDB}, the scalar curvature functional reads
 \begin{align*}
 \mathscr{R}_D(f)&=\frac{2^{n-3}}{3}(n-2)\nu_{n-1}\int_M \dv f\left[-R-3T^0_{baa} T^0_{bcc}
 +\frac{3}{4}\bigl(T^0_{abc}T^0_{abc}+2T^0_{abc}T^0_{cab}\bigr)\right]\\
 &=\frac{2^{n-3}}{3}(n-2)\nu_{n-1}\int_M \dv f\left[-R-3T^0_{baa}T^0_{bcc}+\frac{9}{4}{}^A T^{0}_{abc} {}^A T^{0}_{abc}\right].
 \end{align*}
\end{Proposition}
\begin{proof}
 From \cite[Lemma~2.1]{dabrowski2024a}, we immediately infer that
 \[
 \Tr\bigl(P^0_{aa}\bigr)=\frac{2^n}{3}R, \qquad \Tr\bigl(Q^0\bigr)=\frac{2^{n-2}}{3}R.
 \]
 Combining this with Lemma~\ref{lemma:TRB} and Proposition~\ref{prop:SA}, the result follows.
\end{proof}

\begin{Remark}
 We observe that if the vector part of the torsion vanishes, the scalar curvature functional for the Hodge--Dirac operator in Proposition~\ref{prop:HD_SA} aligns with the scalar curvature functional for the spin Dirac operator from Theorem~\ref{EH-spinD}, up to an overall coefficient. We can summarize the result for the effective density of the functional scalar curvature in the presence of torsion, $R_T$, as
 \[
 R_T=R-
 \frac{9}{4}T^{A0}_{abc}T^{A0}_{abc}+\begin{cases}
 0,& \mbox{spin Dirac},\\
 3T^0_{baa}T^0_{bcc}, & \mbox{Hodge--Dirac}.
 \end{cases}
 \]
Note that for the vector torsion, which is permitted in the case of the Hodge--Dirac operator, the sign differs from that of the antisymmetric part of the torsion, and thus the functional may not have extrema.
\end{Remark}

\section{Chiral spectral functionals}
An interesting generalization of the spectral functionals defined so far appears when one includes the grading $\chi$
for even spectral triples. We refer to them as \emph{chiral} spectral functionals.
Their geometrical meaning is rather simple, as they provide
pseudoscalar quantities instead of scalar ones. This extends the realm of
possible functionals and, as we shall discuss in this section for the spin manifolds, allows us to represent some obvious pseudoscalar functionals as
chiral spectral functionals. It remains to be studied in more detail whether
some functionals of this type can capture, for example, other perturbations
of the Dirac operator that arise from torsion.
\begin{Definition}
For an even spectral triple with the Dirac operator $D$, the chiral metric, Einstein, torsion, and scalar curvature functionals are, respectively, defined by
\begin{gather*}
\mathscr{g}^\chi_D(u,w)=\W(\chi \hat{u}\hat{w}|D|^{-n}), \qquad u,w\in \Omega_D^1, \\
	\mathscr{G}_D^\chi(u,w)=\W\bigl(\chi\hat{u}\{D,\hat{w}\} D |D|^{-n}\bigr), \qquad u,w\in \Omega_D^1, \\
	\mathscr{T}_D^\chi (u,v,w)=\W\bigl(\chi\hat{u}\hat{v}\hat{w}D D^{-2m}\bigr), \qquad u,v,w\in \Omega^1_D,
\end{gather*}
and
\[
 \mathscr{R}_D^\chi(f)=\W\bigl(\chi f D^{-2m+2}\bigr), \qquad f\in \mathcal{A}.
\]
\end{Definition}

In the following, we assume that the Dirac operator $D$ decomposes as $D=D_0+B$, where~$D_0$ satisfies the assumptions outlined in the previous section and anticommutes with the grading~$\chi$. Additionally, we require that the perturbation $B$ anticommutes with $\chi$ to ensure the anticommutativity of the full Dirac operator with this chiral structure. First, note that, similar to the non-chiral case, for $D=D_0+B$ satisfying the assumptions of this paper, the chiral metric functional satisfies $\mathscr{g}^\chi_D(u,w) = \mathscr{g}^\chi_{D_0}(u,w)$. Next, we have the following.

\begin{Proposition}
	\label{prop:Einchi}
	For $D = D_0+B$ satisfying the assumption of this paper, the chiral Einstein functional reads
		\begin{align*}
			\mathscr{G}_D^\chi(u,w)={}& \mathscr{G}_{D_0}^\chi(u,w)
 + \nu_{n-1}\int_M\Tr \chi\Bigr\{
 {\rm i}u_a w_{bc}\bigl[2(3-n)\gamma^a \delta^{bc}+2\delta^{ab} \gamma^c -2\delta^{ac} \gamma^b \\
 &-(4-n)\gamma^a\gamma^b\gamma^c\bigr]B_0+{\rm i}u_a w_b \bigl[(3-n)\gamma^a B_b -\gamma^b B_a +\delta^{ab}\gamma^c B_c - \gamma^a\gamma^b\gamma^c B_c\bigr]\\
 & +\frac{1}{4}u_aw_b \bigl(\bigl[\gamma^b,\gamma^a\bigr] B_0 -2\gamma^a B_0 \gamma^b\bigr)\gamma^c B_0 \gamma^c
 \Bigr\}.
		\end{align*}
\end{Proposition}
\begin{proof}
 The reasoning is analogous to the one in the proof of \cite[Proposition~II.1.]{bochniak2025}, and we get
\begin{gather*}
				\mathscr{G}_D^\chi(u,w)-\mathscr{G}_{D_0}^\chi(u,w) \\
			\qquad=\frac{\nu_{n-1}}{24}\int\operatorname{Tr}\bigl(24\chi u_a\gamma^a\bigl( {\rm i}\gamma^c\gamma^bw_{bc}B_0 + {\rm i} \gamma^c\gamma^bw_bB_c+w_bB_0\gamma^bB_0\bigr)\nonumber\\
			 \phantom{\qquad=}{} +12 {\rm i} \chi\bigl[-u_aw_{bd}\gamma^a\gamma^d\gamma^b\gamma^c
			 + {\rm i}u_aw_b\gamma^a(2\delta^{bc}B_0
			 + \bigl\{\gamma^b,B_0\bigr\}\gamma^c - \gamma^b\{\gamma^c,B_0\})\bigr]\{\gamma^c,B_0\}\\
			 \phantom{\qquad=}{} +6\chi\bigl(u_cw_b\gamma^c\gamma^a + u_cw_a\gamma^c\gamma^b
			 - u_cw_d\gamma^c\gamma^d\delta_{ab}\bigr)\bigl[-2 {\rm i}\{\gamma^a,B_b\} + \bigl\{\gamma^a,B_0\bigr\}\bigl\{\gamma^b,B_0\bigr\}\bigr]\bigr).
		\end{gather*}
	
	First, we collect the terms with $w_{bc}$
		\begin{gather*}
			\frac{\rm i}{2}u_aw_{bc}\Tr \chi \bigl[2\gamma^a\gamma^c\gamma^bB_0-\gamma^a\gamma^c\gamma^b\gamma^d\bigl\{\gamma^d,B_0\bigr\}\bigr]\\
			\qquad=\frac{\rm i}{2}u_aw_{bc}\operatorname{Tr} \chi \bigl[(2-n)\gamma^a\gamma^c\gamma^bB_0-\gamma^a\gamma^c\gamma^b\gamma^dB_0\gamma^d\bigr]
			\\
			\qquad=\frac{\rm i}{2}u_aw_{bc}\Tr \bigl[(2-n)\chi \gamma^a\gamma^c\gamma^bB_0
			+ 2\delta^{ad}\chi \gamma^c\gamma^b\gamma^dB_0
			- 2\delta^{cd}\chi \gamma^a\gamma^b\gamma^dB_0\\
		\phantom{\qquad=}{}	+ 2\delta^{bd}\chi \gamma^a\gamma^c\gamma^dB_0
			-n \chi \gamma^a\gamma^c\gamma^bB_0\bigr]
			\\
			\qquad= {\rm i} u_a w_{bc}\Tr \bigl[(2-n) \chi \gamma^a \gamma^c\gamma^b B_0 +\chi \gamma^c \gamma^b \gamma^a B_0-\chi \gamma^a \gamma^b \gamma^c B_0\bigr]\\
			\qquad= {\rm i}u_a w_{bc} \Tr\bigl[(4-n) \chi\gamma^a \gamma^c \gamma^b B_0 +2\chi\bigl(\delta^{ab}\gamma^c -\delta^{ac} \gamma^b -\delta^{bc}\gamma^a\bigr)B_0\bigr]\\
			\qquad= {\rm i}u_a w_{bc}\Tr\bigl[\chi\bigl((6-2n)\gamma^a \delta^{bc}+2\delta^{ab}\gamma^c -2\delta^{ac}\gamma^b-(4-n)\gamma^a\gamma^b \gamma^c \bigr)B_0\bigr].
		\end{gather*}
	Next, we collect the remaining terms that are linear in $B$ into
		\begin{gather*}
			\frac{\rm i}{2}u_a w_{b} \Tr \bigl[
			\chi\gamma^a \bigl(2\gamma^c \gamma^b B_c -\gamma^c\{\gamma^c, B_b\}-\gamma^c \{\gamma^b, B_c\}+\gamma^b \{\gamma^c,B_c\}\bigr) \bigr] \\
			\qquad=\frac{\rm i}{2} u_a w_b \Tr \bigl[
			\chi\gamma^a \bigl(\gamma^c \gamma^b B_c -n B_b -\gamma^c B_b \gamma^c -\gamma^c B_c \gamma^b +\gamma^b \gamma^c B_c +\gamma^b B_c \gamma^c\bigr)
			\bigr] \\
			\qquad=\frac{\rm i}{2}u_a w_b \Tr \bigl[
			\chi\gamma^a \bigl\{\gamma^c,\gamma^b\bigr\} B_c -n\gamma\gamma^a B_b -\gamma\gamma^a \gamma^c B_b \gamma^c -\gamma \gamma^a \gamma^c B_c \gamma^b +\gamma\gamma^a\gamma^b B_c \gamma^c
			\bigr] \\
			\qquad=\frac{\rm i}{2} u_a w_{b}\Tr \bigl[(2-n) \chi\gamma^a B_b +\gamma (2\delta^{ac} -\gamma^a\gamma^c)\gamma^c B_b +\gamma\gamma^b\gamma^a \gamma^c B_c +\gamma\gamma^a\gamma^b B_b \gamma^c\bigr] \\
			\qquad=\frac{\rm i}{2} u_a w_b \Tr \bigl[
			(2-n)\chi\gamma^a B_b +2\gamma \gamma^a B_b -n\gamma \gamma^a B_b +\gamma\gamma^b \gamma^a \gamma^c B_c -\gamma\gamma^c \gamma^a \gamma^b B_c\bigr] \\
			\qquad=\frac{\rm i}{2} u_a w_b \Tr \bigl[
			(2-n) \chi\gamma^a B_b +2\gamma\gamma^a B_b -n \gamma\gamma^a B_b +\gamma\gamma^b \gamma^a\gamma^c B_c -\gamma (2\delta^{ac}-\gamma^a \gamma^c)\gamma^b B_c \bigr] \\
			\qquad=\frac{\rm i}{2}u_a w_b \Tr \bigl[
			2(2-n)\chi\gamma^a B_b +\gamma\gamma^b \gamma^a \gamma^c B_c -2\gamma\gamma^b B_a +\gamma\gamma^a \bigl(2\delta^{cb}-\gamma^b \gamma^c\bigr)B_c
			\bigr] \\
			\qquad=\frac{\rm i}{2} u_a w_b \Tr \bigl[2(2-n)\chi\gamma^a B_b +\gamma\gamma^b \gamma^a \gamma^c B_c -2\gamma\gamma^b B_a +2\gamma\gamma^a B_b -\gamma\gamma^a\gamma^b \gamma^c B_c\bigr] \\
			\qquad=\frac{\rm i}{2}u_a w_b \Tr \bigl[
			2(3-n)\chi\gamma^a B_b +\gamma \bigl(2\delta^{ab}-\gamma^a\gamma^b\bigr)\gamma^c B_c -2\gamma\gamma^b B_a -\gamma \gamma^a\gamma^b \gamma^c B_c\bigr] \\
			\qquad= {\rm i}u_a w_b\Tr \bigl[(3-n)\chi\gamma^a B_b -\gamma\gamma^b B_a +\delta^{ab}\gamma\gamma^c B_c -\gamma\gamma^a \gamma^b \gamma^c B_c\bigr].
		\end{gather*}
 Finally, the terms quadratic in $B$ are
 \begin{gather*}
			\frac{1}{4}u_aw_b\Tr \chi\gamma^a\bigl[4B_0\gamma^bB_0-4B_0\bigl\{\gamma^b,B_0\bigr\}-2\bigl\{\gamma^b,B_0\bigr\}\gamma^c\{\gamma^c,B_0\}+2\gamma^b\{\gamma^c,B_0\}^2\\
		\qquad+\gamma^c\{\gamma^c,B_0\}\bigl\{\gamma^b,B_0\bigr\}+\gamma
 ^c\bigl\{\gamma^b,B_0\bigr\}\{\gamma^c,B_0\}-\gamma^b\{\gamma^c,B_0\}^2\bigr]\\
		\phantom{\qquad+}{}	=\frac{1}{4}u_aw_b \Tr \chi \gamma^a\bigl[
 4 B_0 \gamma^b B_0 -4 B_0 \gamma^b B_0 -4 B_0^2 \gamma^b +\gamma^c\{\gamma^c, B_0\}\gamma^b B_0\\
\phantom{\qquad+=}{} +\gamma^c\{\gamma^c, B_0\}B_0 \gamma^b	+\bigl(-2\gamma^b B_0 \gamma^c -2B_0 \gamma^b \gamma^c +2\gamma^b \gamma^c B_0 +2\gamma^b B_0 \gamma^c \\
\phantom{\qquad+=}{}	+\gamma^c \gamma^b B_0 +\gamma^c B_0 \gamma^b-\gamma^b\gamma^c B_0 -\gamma^b B_0 \gamma^c\bigr)\{\gamma^c, B_0\}
 \bigr]\\
 \phantom{\qquad+}{}	=\frac{1}{4}u_aw_b \Tr \chi\bigl(4\gamma^b \gamma^a B_0^2 +n\gamma^a B_0 \gamma^b B_0+\gamma^a \gamma^c B_0 \gamma^c \gamma^b B_0 -n\gamma^b \gamma^a B_0^2\\
\phantom{\qquad+=}{}	-\gamma^b \gamma^a \gamma^c B_0 \gamma^c B_0+2\gamma^a B_0 \bigl\{\gamma^b, B_0\bigr\}-2n \gamma^a B_0 \gamma^b B_0 -2\gamma^a B_0 \gamma^b \gamma^c B_0 \gamma^c \\
\phantom{\qquad+=}{}+\gamma^a \gamma^c B_0 \gamma^b \gamma^c B_0+\gamma^a \gamma^c B_0 \gamma^b B_0 \gamma^c - n\gamma^a \gamma^b B_0^2 -\gamma^a\gamma^b B_0 \gamma^c B_0 \gamma^c\bigr)\\
\phantom{\qquad+}{} =\frac{1}{4}u_aw_b \Tr \chi\bigl[
 (2-n)\gamma^a\gamma^b B_0^2 +(2-n)\gamma^b\gamma^a B_0^2 +(2-n) \gamma^a B_0\gamma^b B_0 \\
\phantom{\qquad+=}{}-\gamma^b\gamma^a \gamma^c B_0 \gamma^c B_0+4B_0 \gamma^b \gamma^a B_0-2\gamma^a\gamma^c B_0\gamma^b \gamma^c B_0 \\
\phantom{\qquad+=}{}+(n-2)\gamma^a B_0 \gamma^b B_0 - \gamma^a \gamma^b B_0 \gamma^c B_0 \gamma^c
 \bigr]\\
 \phantom{\qquad+}{}=\frac{1}{4}u_a w_b \Tr \chi\bigl[
2(2-n) \delta^{ab}B_0^2 -2\gamma^b B_0 \gamma^a B_0 +2\gamma^a B_0 \gamma^b B_0 +\gamma^b \gamma^a B_0 \gamma^c B_0 \gamma^c \\
\phantom{\qquad+=}{} + 4 B_0 \gamma^b \gamma^a B_0-4 B_0 \gamma^b \gamma^a B_0 -2\gamma^a B_0 \gamma^b \gamma^c B_0 \gamma^c
 -\gamma^a \gamma^b B_0 \gamma^c B_0 \gamma^c\bigr]\\
 \phantom{\qquad+}{}=\frac{1}{4}u_q w_b \Tr \chi \bigl(\bigl[\gamma^b,\gamma^a\bigr]B_0 - 2\gamma^a B_0 \gamma^b\bigr) \gamma^c B_0 \gamma^c.	\tag*{\qed}
 \end{gather*} \renewcommand{\qed}{}
\end{proof}

\begin{Proposition}
	\label{prop:torchi}
 For $D = D_0+B$ satisfying the assumption of this paper, the chiral torsion functional reads
 \[
 \mathscr{T}^\chi_D(u,v,w)=\frac{\nu_{n-1}}{2}\int_M \dv \Tr\left[\chi\hat{u}\hat{v}\hat{w}(2B_0-\gamma^a\{\gamma^a,B_0\})\right].
 \]
\end{Proposition}
\begin{proof}
 It follows from Proposition~\ref{lemma:ED}.
\end{proof}

\begin{Proposition}
	\label{prop:EHchi}
 For $D = D_0+B$ satisfying the assumption of this paper, the chiral scalar curvature functional reads
 \[
 \mathscr{R}^\chi_D(f)=\mathscr{R}^\chi_{D_0}(f)- {\rm i} \frac{\nu_{n-1}}{2}(n-2) \int_M \dv \Tr (\chi \gamma^a B_a).
 \]
 Since $B_a=\partial_a B$, the second term is a boundary term and as such vanishes on closed manifolds.
\end{Proposition}
\begin{proof}
By Corollary~\ref{lemma:E}, we have
\[
 \W\bigl(\chi D^{-n+2}\bigr)=\frac{\nu_{n-1}}{24}(n-2)\int_M \dv \Tr \bigl[\chi(-12Q -2R +6P_{aa}-3S_aS_a)\bigr],
\]
so that
\begin{gather*}
 \W\bigl(\chi D^{-n+2}\bigr)-\W\bigl(\chi D_0^{-n+2}\bigr)\\
\qquad =\frac{\nu_{n-1}}{24}(n-2)\int_M \dv \Tr \bigl[\chi\bigl(-12 B_0^2-12{\rm i}\gamma^a B_a +6{\rm i}\{\gamma^a, B_a\}+3\{\gamma^a, B\}\{\gamma^a, B\}\bigr)\bigr]\\
\qquad=\frac{\nu_{n-1}}{8}(n-2)\int_M \dv \Tr \bigl[\chi\bigl(-4B_0^2 -4{\rm i} \gamma^a B_a +\{\gamma^a, B\}\{\gamma^a, B\}\bigr)\bigr]\\
 \qquad=-\frac{\rm i}{2}(n-2)\nu_{n-1}\int_M\dv \Tr(\chi \gamma^a B_a).\tag*{\qed}
\end{gather*} \renewcommand{\qed}{}
\end{proof}

\subsection{The spin Dirac operator}
First, recall that for the spin-Dirac operator, the grading $\chi$ is the chirality operator $\gamma$ in the associated Clifford algebra.
\begin{Proposition}\label{ch-g-spin}
	For the torsion-less spin-Dirac operator $D_0$, the functional $\mathscr{g}_{D_0}^\chi(u,w)$ reads
 \[
 \mathscr{g}_{D_0}^\chi(u,w)=
 \begin{cases}
 \displaystyle 4\pi {\rm i}{2^m}\int_M\dv u_a w_b \varepsilon^{ab}, & n=2,\\
 0, & n>2.
 \end{cases}
 \]
\end{Proposition}

\begin{proof}
 It follows from Proposition~\ref{prop:main}.
\end{proof}

\begin{Proposition}
	For the spin-Dirac operator with torsion, $D=D_0+B$ with $B$ given by \eqref{eq:B}, the functional $\mathscr{T}_{D}^\chi(u,v,w)$ reads
 \[
 \mathscr{T}_{D}^\chi(u,v,w)=2^m \cdot \begin{cases}
 0, & n=2 \text{ or } n\geq 8,\\
 \displaystyle -{\rm i} \frac{\pi^2}{2} \int_M \dv u_a v_b w_c &\\
 \qquad {} \times {}^AT_{ijk} \varepsilon_{ijkl}
 \bigl(\delta_{ab}\delta_{cl} + \delta_{al}\delta_{bc}-\delta_{ac}\delta_{bl}\bigr), & n=4,\\
\displaystyle \frac{\pi^3}{4}\int_M\dv u_av_b w_c {}^AT_{ijk} \varepsilon_{abcijk}, & n=6.
 \end{cases}
 \]
\end{Proposition}
\begin{proof}
First, by \eqref{eq:slady_B_spinowe}, we have
\[
 \mathscr{T}^\chi_D(u,v,w)=\frac{{\rm i}\nu_{n-1}}{4}\int_M \dv u_a v_b w_c {}^AT_{ijk} \Tr\bigl(\gamma \gamma^a\gamma^b\gamma^c\gamma^i\gamma^j\gamma^k\bigr).
\]
Notice that for either $n=2$ or $n\geq 8$ (recall that here $n=2m$), this expression vanishes identically. Therefore, the only potentially non-zero contribution to the torsion functional could arise in dimension $4$ or $6$. For $n=4$, the torsion functional reads
\begin{align*}
 \mathscr{T}^\chi_D(u,v,w)&=-{\rm i}\frac{\nu_3}{4}\int_M \dv u_a v_b w_c {}^AT_{ijk} \varepsilon_{ijkl} \Tr\bigl(\gamma^a\gamma^b \gamma^c \gamma^l\bigr)
 \\
 &=-\frac{{\rm i}\nu_3}{4} \cdot 2^m\int_M \dv u_a v_b w_c {}^AT_{ijk} \varepsilon_{ijkl}
 (\delta_{ab}\delta_{cl} + \delta_{al}\delta_{bc}-\delta_{ac}\delta_{bl}).
 \end{align*}
On the other hand, for $n=6$ we immediately obtain
\[
 \mathscr{T}^\chi_D(u,v,w)=\frac{\nu_5}{4}\cdot 2^m\int_M\dv u_av_b w_c {}^AT_{ijk} \varepsilon_{abcijk}.\tag*{\qed}
\] \renewcommand{\qed}{}
\end{proof}

\begin{Proposition}
	For the torsion-less spin-Dirac operator $D_0$, the functional $\mathscr{G}_{D_0}^\chi(u,w)$ vanishes.
\end{Proposition}
\begin{proof}
	First, we write the operator $\gamma\hat{u}\{D_0,\hat{w}\}D_0|D_0|^{-n}$ as $(O_1+O_2)L^{-m}$, where
	\[
		O_1=\gamma\hat{u}D_0\hat{w} D_0, \qquad O_2=\gamma \hat{u}\hat{w}D_0^{2}.
	\]
	With the notation from Section~\ref{sec:general}, we have $P_{ab}=\frac{2}{3}\mathrm{Ric}_{ab}+\frac{1}{4}R_{abjk}\gamma^j\gamma^k$, $S_a=0$ and $Q=\frac{1}{4}R$. For the operator $O_1$, we have
	\begin{gather*}
			F^{jk}=u_a w_b \gamma \gamma^a \gamma^{(j|} \gamma^b \gamma^{|k)}= u_a\bigl(w_k\gamma \gamma^a \gamma^j +w_j\gamma\gamma^a\gamma^k -w_b \gamma\gamma^a\gamma^b \delta^{jk}\bigr),\\
			G^k=-u_aw_{bj}\gamma \gamma^a \gamma^j \gamma^b \gamma^k,\\
			H=-\frac{1}{8}u_a w_b \gamma \gamma^a \gamma^j \gamma^b \gamma^k \gamma^p\gamma^s R_{jkps}=\frac{1}{4}u_aw_b \gamma\gamma^a \gamma^j \gamma^b\gamma^d \mathrm{Ric}_{jd}\\
		\phantom{H}{}	=\frac{1}{2}u_aw_b \mathrm{Ric}_{jb}\gamma\gamma^a\gamma^j -\frac{1}{4}u_aw_b \mathrm{Ric}_{jd}\gamma\gamma^a \gamma^{[j}\gamma^{d]}\gamma^b	=\frac{1}{2}u_aw_b \mathrm{Ric}_{jb}\gamma\gamma^a\gamma^j -\frac{1}{4}u_a w_b R \gamma\gamma^a \gamma^b.
		\end{gather*}
	We notice that, for $n\neq 2$, $O_2=-\frac{1}{n-2}F^{aa} L$, and therefore, Corollary~\ref{cor4} gives us
	\begin{align*}
			\mathscr{G}_{D_0}^\chi(u,w)&=\frac{\nu_{n-1}}{24}\int_M\dv \Tr \bigl(24H -12F^{ab}P_{ab}+4F^{ab}\mathrm{Ric}_{ab}\bigr)\\
			&=\frac{\nu_{n-1}}{6}\int_M\dv \Tr\bigl(6H -F^{ab}\mathrm{Ric}_{ab}\bigr).
		\end{align*}
	Since $\Tr\bigl(\gamma\gamma^a\gamma^b\bigr)=2i\delta_{n,2}\epsilon^{ab}$, from the form of the $F^{jk}$ and $H$ operators, we infer that $\mathscr{G}_{D_0}^\chi(u,w)$ vanishes identically for $n\neq 2$. For $n=2$, we are, strictly speaking, not allowed to use Corollary~\ref{cor4} directly, but we refer instead to Proposition~\ref{prop:main} applied to $O=O_1$ and Theorem~\ref{lemma:E} applied to $E=O_2 L^{-1}$. From the latter, we get $\W\bigl(EL^{-m+1}\bigr)=0$, while the former, due to the fact that in $n=2$ dimensions we have $F^{jj}=0$, again leads to
	\[
			\mathscr{G}_{D_0}^\chi(u,w)=\frac{\nu_{n-1}}{6}\int_M\dv \Tr \bigl(6H -F^{ab}\mathrm{Ric}_{ab}\bigr),
		\]
	so that, in this case,
	\begin{align*}
			\mathscr{G}_{D_0}^\chi(u,w)={}&{\rm i}\frac{\nu_{1}}{3}\int_M\dv \biggl(3u_aw_b \mathrm{Ric}_{jb}\epsilon^{aj}-\frac{3}{2}u_aw_b R \epsilon^{ab}-u_a w_k \epsilon^{aj}\mathrm{Ric}_{jk} \\
			& -u_a w_j \epsilon^{ak}\mathrm{Ric}_{jk}+u_aw_b \epsilon^{ab}\delta^{jk}\mathrm{Ric}_{jk}\biggr)\\
			={}&{\rm i}\frac{\nu_{1}}{6}\int_M\dv\bigl[\mathrm{2Ric}_{jb}\epsilon^{aj}-\epsilon^{ab}R\bigr]=0,
		\end{align*}
	since for two-dimensional manifolds it is known that $\mathrm{Ric}=\frac{R}{2}g$.
\end{proof}

\begin{Proposition}
 For the spin-Dirac operator with torsion, the functional $\mathscr{G}_{D}^\chi(u,w)$ reads
 \[
 \mathscr{G}_{D}^\chi(u,w)=\begin{cases}
\displaystyle 2\pi^2 \int_M \dv u_a \biggl[T_{ijk}^0 (\varepsilon_{ijka} w_{bb}-\varepsilon_{ijkc} w_{ac} +\varepsilon_{ijkb}w_{ba}) & \\
\displaystyle \qquad{}+ w_b \varepsilon_{ijka}T_{ijk}^c +\frac{3}{4}w_b \varepsilon_{ijkl}T^0_{ijk}T^0_{abl}\biggr], & n=4,\\
\displaystyle {\rm i}\pi^3 \int_M \dv u_a \bigl(2 w_{bc} T^0_{ijk}\varepsilon_{ijkabc} -w_b T^c_{ijk} \varepsilon_{ijkabc}\bigr), & n=6,\\
 0, & \text{otherwise}.
 \end{cases}
 \]
\end{Proposition}
\begin{proof}
First, by counting the number of gamma matrices, we notice that for any $n\neq 4,6$, the chiral Einstein functional vanishes.

For $n=4$, we have $\gamma B=-B\gamma =-\frac{\rm i}{8}\varepsilon_{abcd}T_{abc}\gamma^d$, and the claim follows by a straightforward computation using Proposition~\ref{prop:Einchi}. For $n=6$ we have $B\gamma=-\gamma B=-\frac{1}{48}T_{abc} \varepsilon_{abcijk}\gamma^i\gamma^j\gamma^k$ and~$\bigl\{\gamma^d,\varepsilon_{abcijk}\gamma^i\gamma^j\gamma^k\bigr\}=6\varepsilon_{abcdjk}\gamma^j\gamma^k$, so that
\smash{$
 \Tr\bigl(\gamma^d\gamma^e\gamma^f \varepsilon_{abcijk}\gamma^i\gamma^j\gamma^k\bigr)=-6\varepsilon_{abcdef}$}.
This, together with Proposition~\ref{prop:Einchi} concludes the proof.
\end{proof}

\begin{Proposition}
	For the torsion-less spin-Dirac operator $D_0$, the functional $\mathscr{R}_{D_0}^\chi(f)$ vanishes, and
 \[
 \mathscr{R}_{D}^\chi(f)=-\frac{{2^m}\nu_3}{8}\delta_{n,4}\int_M\dv \varepsilon^{abcd}\partial_a T_{bcd},
 \]
 which also vanishes on closed manifolds.
\end{Proposition}
\begin{proof}
Since in this case we have $B=-\frac{\rm i}{8}T_{abc}\gamma^a\gamma^b \gamma^c$ and $\chi=\gamma$ (the product of all gamma matrices), we have
\begin{gather*}
 -\frac{\rm i}{2}(n-2)\nu_{n-1}\int_M\dv \Tr
 (\chi\gamma^a B_a)\\
 \qquad=-\frac{\rm i}{2}(n-2)\nu_{n-1}\int_M\dv \Tr\biggl[\gamma\gamma^a \biggl(-\frac{\rm i}{8}\partial_a T_{bcd} \gamma^b \gamma^c \gamma^d\biggr)\biggr]\\
 \qquad
 =-\frac{(n-2)\nu_{n-1}}{16}\int_M\dv \Tr\bigl(\gamma\gamma^a \gamma^b \gamma^c\gamma^d\bigr)\partial_a T_{bcd}\\
 \qquad=-{2^m}\frac{\nu_{n-1}(n-2)}{16}\delta_{n,4}\int_M \dv \varepsilon^{abcd} \partial_a T_{bcd}
 =-{2^m}\frac{\nu_3}{8}\delta_{n,4}\int_M \varepsilon^{abcd}\partial_a T_{bcd},
\end{gather*}
and
\[
 \W\bigl(\chi D_0^{-n+2}\bigr)=\frac{\nu_{n-1}}{24}(n-2) \int_M \Tr\bigl[\gamma(-3R-2R+4R)\bigr]=0.\tag*{\qed}
\]\renewcommand{\qed}{}
\end{proof}

\begin{Remark}
 Finally, we note a result similar to the one in Proposition~\ref{rem:1}. Mainly, for $E=u_a\gamma\gamma^a$ in Proposition~\ref{lemma:ED}, we infer that
 \[
\W\bigl(u_a\gamma\gamma^aDD^{-2m}\bigr)={2^m}\frac{{\rm i}\nu_{3}}{4}\delta_{n,4}u_a T_{ijk} \varepsilon_{aijk}.
 \]
\end{Remark}

\subsection{The Hodge--Dirac operator}

The spectral triple of the Hodge--Dirac operator is equipped with two distinct gradings. Before proceeding, we introduce a convenient notation
$
 \gamma^p = -{\rm i}(\lambda_+^p -\lambda_-^p)$, $\tilde{\gamma}^p = \lambda^p_+ +\lambda^p_-$.
Note that these operators are hermitian and satisfy,
\[
\{ \gamma^p, \gamma^s \} = \{\tilde{\gamma}^p, \tilde{\gamma}^s \} = 2 \delta^{ps},
\qquad
\{ \tilde{\gamma}^p, \gamma^s \} = 0.
\]
\subsubsection{The Euler and Hodge gradings}
Here we consider the Hodge spectral triple equipped with a grading given by the form,
\begin{align*}
 &\chi_e= \gamma^1 \gamma^2 \cdots \gamma^n \tilde{\gamma}^1 \tilde{\gamma}^2 \cdots \tilde{\gamma}^n \qquad
 \hbox{(Euler grading)}, \\
 &\chi_h ={\rm i}^m \gamma^1 \gamma^2 \cdots \gamma^n
 \qquad \hbox{(Hodge grading)}.
\end{align*}
It could be easily verified that on a form of degree $p$ on a manifold of dimension $2m$, $\chi_e$ is $(-1)^p$ and $\chi_h$ is simply ${\rm i}^{p(p-1)-m} \ast$, where $\ast$ is the Hodge map. Note that the composition of both gradings is another grading, $\hat{\chi}=\chi_h\chi_e={\rm i}^m \tilde{\gamma}^1 \tilde{\gamma}^2 \cdots \tilde{\gamma}^n$, which, however, commutes with $d+d^\ast$.
Furthermore, note that
$
\chi_h \lambda_\pm^p = \lambda_\mp^p \chi_h$,
$
\hat{\chi} \lambda_\pm^p = -\lambda_\mp^p \hat{\chi}$.

These gradings behave quite differently and, for example, can restrict perturbations of the Dirac operator. While the most general torsion term anticommutes with $\chi_e = \chi_h \hat{\chi}$, we obtain a~nontrivial condition for the anticommutation of the torsion term with $\chi_h$ alone, $T_{ijk} \bigl(\lambda^j_+ \lambda^i_+ \lambda^k_- + \lambda^k_+ \lambda^i_- \lambda^j_-\bigr) = - T_{ijk} \bigl(\lambda^j_- \lambda^i_- \lambda^k_+ + \lambda^k_- \lambda^i_+ \lambda^j_+\bigr)$, leading to $T_{ijk} \bigl(\delta^{ik} \lambda^j_+ - \delta^{jk} \lambda^i_+\bigr) = 0$, so that $T_{ijj}=0$.

We can now proceed with computing the corresponding chiral spectral functionals. Here, we focus only on the chiral metric functional, which is very similar to the case of the spin-Dirac operator (Proposition~\ref{ch-g-spin}).
\begin{Proposition}
	The chiral metric functionals for the Hodge--Dirac operator with torsion, for the gradings $\chi_h$, $\chi_e$, and $\hat{\chi}$, are given by
 \begin{gather*}
 \mathscr{g}_{D_0}^{\chi_h}(u,w)=
 \begin{cases}
 \displaystyle -8\pi {\rm i}\int_M\dv u_a w_b \varepsilon^{ab}, & n=2,\\
 0, & n>2,
 \end{cases}
\qquad
 \mathscr{g}_{D_0}^{\chi_e}(u,w)= 0, \qquad n\geq 2,
 \end{gather*}
 respectively.
\end{Proposition}

It is worth noting that the chiral torsion and Einstein functionals are, in principle, explicitly computable using Propositions~\ref{prop:torchi} and~\ref{prop:Einchi}, respectively. These calculations are not performed in this study, as they do not significantly contribute to the objectives of the current research. Interested readers may pursue the details independently if they wish.

\subsection*{Acknowledgements}
We thank the anonymous referees for their insightful comments. The work of AB was partially funded by the Deutsche Forschungsgemeinschaft (DFG, German Research Foundation) under Germany's Excellence Strategy -- EXC-2111 -- 390814868. AB is supported by the Alexander von Humboldt Foundation. A.S.\ is supported by the Polish National Science Centre grant 2020/37/B/ST1/01540. L.D.\ is affiliated with GNFM–INDAM (Istituto Nazionale di Alta Matematica) and acknowledges that this research is part of the EU Staff Exchange project 101086394 ``Operator Algebras That One Can See''.

\pdfbookmark[1]{References}{ref}
\LastPageEnding

\end{document}